\documentclass[twocolumn,notitlepage,pra,showpacs,superscriptaddress,amssymb,amsmath,amsmath]{revtex4-2}
\usepackage{lipsum}
\usepackage{subcaption} 
\usepackage{graphicx}
\usepackage{epstopdf}
\usepackage{dcolumn}
\usepackage{bm}
\usepackage{physics}
\usepackage{xcolor}
\usepackage{amssymb}
\usepackage{gensymb}
\usepackage{amsmath}
\usepackage{bbold}
\usepackage{soul}
\usepackage[colorlinks=true, linkcolor=blue, urlcolor=blue, citecolor=blue]{hyperref}

\usepackage[T1]{fontenc} 

\newcommand{\affFUW}{Faculty of Physics, University of Warsaw, Pasteura 5, 02-093 Warsaw, Poland}
\newcommand{\affCFT}{Center for Theoretical Physics, Polish Academy of Sciences, Al. Lotników 32/46, 02-668 Warsaw, Poland}

\preprint{APS/123-QED}

\begin{document}

\title{Effective Interactions in Quasi-One-Dimensional Dipolar Quantum Gases}

\author{Michał Zdziennicki}
  \affiliation{\affFUW}
 \author{Mateusz Ślusarczyk}%
 \affiliation{\affCFT}%
 \author{Krzysztof Pawłowski}%
 \affiliation{\affCFT}%
 \author{Krzysztof Jachymski}
  \affiliation{\affFUW}

\date{\today}

\begin{abstract}
Ultracold dipolar atoms and molecules provide a flexible quantum simulation platform for studying strongly interacting many-body systems. Determining microscopic Hamiltonian parameters of the simulator is crucial for it to be useful. We study effective interactions emerging in quasi-one-dimensional (q1D) dipolar quantum gases, revealing significant nonuniversal corrections to the commonly used 1D pseudopotential. We demonstrate that a full 3D treatment employing realistic interaction potentials is essential for describing the reduced-dimensional system. Our findings are particularly relevant to experiments probing excited states and nonequilibrium phenomena.
\end{abstract}

\maketitle


\section{Introduction} 
Quasi-one-dimensional (q1D) quantum gases remain an active field of research, driven by unique phenomena arising from reduced dimensionality~\cite{Giamarchi}. The field was driven by mathematical results,  with key findings of E.~Lieb~\cite{Lieb1963May,Lieb1963Excited},  M. Girardeau \cite{Girardeau1960Nov}, and McGuire~\cite{McGuire1964May} showing that strictly 1D Bose gas with short-range interactions remains stable in all interaction regimes.
One-dimensional system with short-range interaction can be implemented in a 3D ultracold dilute gases confined in an elongated, cigar-like harmonic trap
\cite{Lieb1998, Astrakharchik2002, Lieb2003, Seiringer2008Dec}, 
which has been extensively realized experimentally~\cite{Greiner2001Oct, Moritz2003Dec, Tolra2004May, Esteve2006Apr, Kinoshita2006Apr, Hofferberth2007Sep, Haller2009Sep}, and used to demonstrate broad range of phenomena, such as prethermalization ~\cite{Gring2012Sep, Langen2016Jun},  solitons~\cite{Khaykovich2002May, Strecker2002May,Burger1999}, reduction of atomic losses~\cite{Tolra2004May}, or super-Tonks-Girardeau gas~\cite{Haller2010Apr}.

Recently in the field of ultracold gases a lot of attention is devoted to systems with long-range interactions. Shortly after achieving Bose-Einstein condensate of highly magnetic atoms, Erbium and Dysprosium, \cite{Aikawa2012, Mingwu2011}, the 3D quantum droplets have been discovered~\cite{Kadau_2016, Chomaz2016, Schmitt2016Nov} which allowed the observation of supersolidity in cigar-shaped traps \cite{Guo2019Oct, Chomaz2019}. In parallel, the dipolar q1D bosons attracted theoretical attention~\cite{arkhipov_ground-state_2005, Girardeau2012Dec, Parker2008, Bland2015, Pawlowski2015Oct,  Baizakov2009Aug, Corson2013, Oldziejewski2020, Edmonds2020Dec,dePalo2022,DePalo2020Jan, Pal2022, Oldziejewski2018Dec, Kristinsdottir2013, Zin2021Sep},
due to the possibility to generate the roton-maxon excitation spectrum \cite{Corson2013, Oldziejewski2018Dec}, the supersolid and the quantum self-bound states \cite{arkhipov_ground-state_2005, Edmonds2020Dec, Baizakov2009Aug, Oldziejewski2020, dePalo2022, Morera2022b, Morera2023, Kraus2022, Bland2015, Pawlowski2015Oct, Bland2017, Astrakharchik2008, Marciniak2023, Girardeau2012Dec, Molignini2025}, dipolar bubbles \cite{Edmonds2020Dec}, different phases of matter \cite{Morera2022b, Morera2023, Kraus2022} and anomalous solitons \cite{Bland2015, Pawlowski2015Oct, Bland2017}, Tonks-Girardeau states \cite{arkhipov_ground-state_2005, Astrakharchik2008, Marciniak2023}, unusual super-Tonks-Girardeau effect \cite{Girardeau2012Dec, Marciniak2023, Molignini2025}. 
The q1D dipolar gases are already under the experimental study ~\cite{Kao2021Jan,Yang2023Aug, Li2023Jun} showing their enhanced stability that allowed for topological pumping to bring the system to stable excited quantum states~\cite{Kao2021Jan, Yang2023Aug, Li2023Jun}. They are also promising candidates for quantum simulation of extended Hubbard models~\cite{Su2023,Du2024}.

Accurate description of the dipolar quasi-1D system  in terms of a purely 1D Hamiltonian is a subtle problem. The quasi-1D regime is typically achieved by confining the gas in a tight harmonic trap, constraining the atomic motion in two transverse directions. In such geometry, the 3D interaction between dipolar atoms containing the short-range $V_{\rm sr}(\bm{r})$ and dipole-dipole $V_{\rm dd}(\bm{r})$ parts with strengths $g^{({\rm 3D})}$ and $g^{({\rm 3D})}_{\rm dd}$ respectively,
\begin{equation}
    V^{({\rm 3D})}(\bm{r}) = g^{({\rm 3D})} V_{\rm sr}(\bm{r}) + g^{({\rm 3D})}_{\rm dd}V_{\rm dd}(\bm{r}),
\end{equation}
has to be reduced to a 1D interaction potential
\begin{equation}
    U^{(\rm 1D)}(x) = g^{\rm (1D)} \,\delta(x) + g^{\rm (1D)}_{\rm dd}U_{\rm dip}(x)\, ,
    \label{eq:effect-dip-1D}
\end{equation}
which also contains a short-range 1D part (resulting from $g^{({\rm 3D})}$ and $g^{({\rm 3D})}_{\rm dd})$)  and a non-local part $U_{\rm dip}(x)$.

\begin{figure*}[!t]
    \centering
    \includegraphics[width=\linewidth]{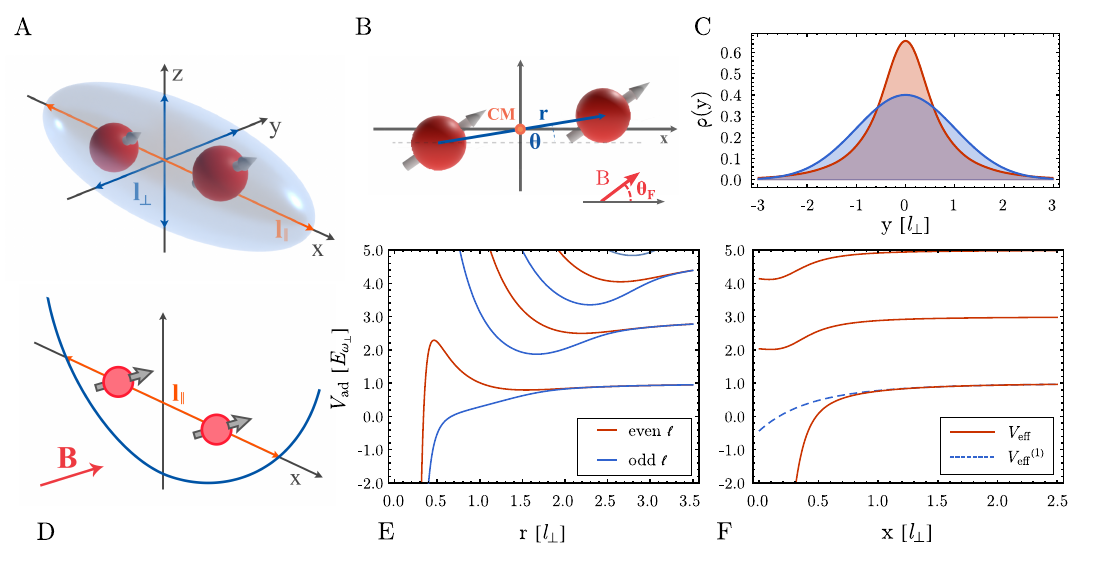}
    \caption{(A) Schematic of the 3D system: Two dipolar particles, aligned by an external magnetic field, are confined in an elongated harmonic trap with transverse size $l_{\perp}$ ($y$, $z$) and longitudinal size $l_{\parallel}$ ($x$). (B) The center-of-mass motion is integrated out, leaving the relative motion with angle $\theta$ in spherical coordinates. (C) Cut through the density profile of the wavefunction minimizing the adiabatic potential \eqref{eq:pot-adiab} at $x=0.5l_{\perp}$ (blue) compared with typically assumed single-particle ground state Gaussian (red).
    (D) Effective 1D system: The 3D system is projected onto 1D, where the dipoles remain aligned by $\mathbf{B}$. (E) Adiabatic energy curves in spherical coordinates obtained for $N=200$ channels. At large separations, the particles approach the harmonic oscillator energy ladder. (F) Same, but in the harmonic oscillator basis; the result for the lowest mode alone is marked by the dashed line. Parameters of the trap are $\omega_{\parallel} = 118 \mathrm{kHz}$ and $\omega_{\perp} = 10\omega_\perp$.}
    
    \label{fig:Rysunek 1}
\end{figure*}

Usually an effective 1D interaction potential \eqref{eq:effect-dip-1D} is obtained assuming that in the transverse degrees of freedom atoms occupy the harmonic oscillator ground state, gaussians, and integrating them out. 
This assumptions is often not fulfilled (see Fig.~\ref{fig:Rysunek 1}C).

The correct way of deriving the appropriate 1D model is by solving the problem of scattering of atoms in the quasi-1D case and expressing the 1D scattering length as a function of 3D parameters. 
In the absence of dipoles, a 1D model with suitable $g^{\rm 1D}$ can be found analytically -- the relevant mapping ~\cite{Olshanii1998Aug, Idziaszek2006} takes the form $g^{\rm (1D)}\propto \frac{A g^{(3D)}}{1- B g^{(3D)}}$ with analytically known coefficients $A$ and $B$.
For the dipolar model, 
finding the mapping is much more difficult, and therefore usually the easy approach is used, assuming single-particle ground state Gaussian in the transverse directions.

It results in a widely used approximate formula for regularized dipole-dipole interaction \cite{ Giovanazzi2004Nov, Deuretzbacher2010, Deuretzbacher2013} and a shift of the $g_{\rm 1D}$~\cite{Deuretzbacher2010, Deuretzbacher2013} (see Eq.~\eqref{Erfc}). Within this approximation, for the angle between the dipole and trap axis $\theta_{\rm F}=\arccos(1/\sqrt{3})\approx 55\degree $, the dipolar contribution vanishes. Experiments~\cite{Lahaye_2009, Pfau2002, Kao2021Jan} have treated this geometry as a reference case with purely contact interaction.
However, at short interatomic distances, the harmonic potential can be neglected and the system is truly three-dimensional. Several theoretical works have already been devoted to solving the scattering problem of dipoles in reduced geometry~\cite{Sinha2007,Giannakeas2013,Shi2014,Guan2014}, using either numerical treatment of model potentials or generalized pseudopotential approach.

In this work, we perform a reliable dimensional reduction using a realistic potential, which consists of van der Waals and dipolar parts, to obtain the 1D parameters $g^{\rm (1D)}$, $g^{\rm (1D)}_{\rm dd}$ and the nonlocal interaction potential $U_{\rm dip}(x)$ as functionals of the original three-dimensional problem. We demonstrate that the long-ranged interaction not only leads to a sizeable energy-dependent correction to the one-dimensional coupling constant $g^{\rm (1D)}$, but due to the coupling to excited transverse states at short-range strong modification of the shape of the effective dipole-dipole potential can be expected. These effects are unavoidable irrespectively from the dipole polarization direction. Our results are vital for detailed understanding of strongly correlated states of dipolar atoms under external confinement and applicable to ongoing experiments~\cite{Tang2018,Kao2021Jan,Korbmacher2023,Su2023}. 

\begin{figure*}
    \centering
    \includegraphics[width=0.9\linewidth]{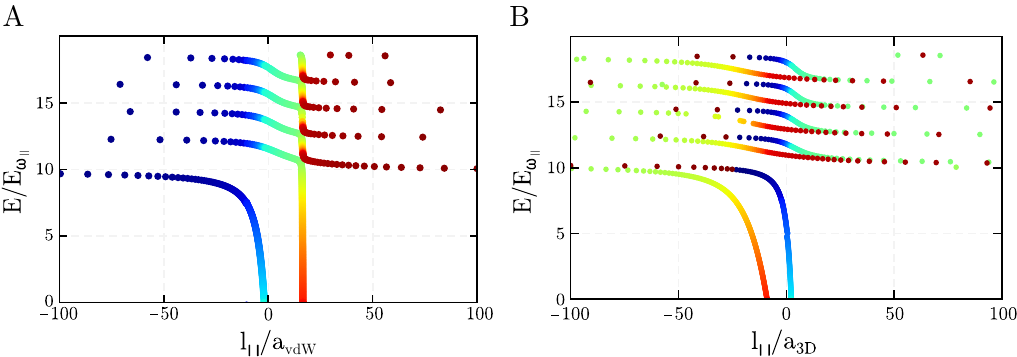}
    \caption{The energy spectrum before and after the conversion from $a_{\text{vdW}}$ (A) to $a_{\text{3D}}$ (B). Respective points are assigned colors based on the order given by the parameter $l_{\text{||}}/a_{\text{vdW}}$. The value is then transformed into $l_{\text{||}}/a_{\text{3D}}$ with color kept. Because of the shifted position of resonance in a harmonic trap the conversion is nonuniversal. The short range $\text{1D}$ solutions are mapped onto two branches.}
    \label{fig:fig2}
\end{figure*}

\section{Hamiltonian and methods}
\subsection{System}
We consider a system consisting of two dipolar atoms confined within a strongly anisotropic ($\omega_{\perp}/\omega_{\parallel} = 10$), cigar-shaped harmonic trap. The magnetic dipole moments are aligned by an external field, forming an angle $\theta_F$ with the elongated axis of the trap, as shown in Fig.~\ref{fig:Rysunek 1}A and \ref{fig:Rysunek 1}B. The atomic interactions are described by both short-range van der Waals and long-range dipole-dipole interactions. The 3D Hamiltonian for the relative distance $\bm{r}$ between  atoms reads:
\begin{equation} 
    \hat{H} = \frac{\hat{p}^2}{2 \mu} + V_{\rm vdW}(r) + V_{\rm dd}(r, \theta) + V_{\rm ho}(\bm{r}), 
    \label{hamiltonian}
\end{equation}
with potential terms:
\begin{align} 
& V_{\rm vdW}(r) = -\frac{C_6}{r^6}, \\
& V_{\rm dd}(r, \theta) = g_{\rm dd}^{(3D)} \frac{1 - 3 \cos^2 \theta}{r^3}, \\
& V_{\rm ho}(\bm{r}) = \frac{1}{2} \mu \big( \omega_{\parallel}^2 z^2 + \omega_{\perp}^2 \rho^2 \big).
\end{align}
 This form of the Hamiltonian is highly relevant in the context of magnetic elements such as Dysprosium and Erbium, although it neglects short-range anisotropies as well as the whole hyperfine structure, which would lead to emergence of multiple channels, possibly with chaotic statistical properties~\cite{Maier2015}. Note that the characteristic interactions lengthscales are comparable with the scale of the transverse trapping potential $l_{\perp} = \sqrt{\hbar / \mu \omega}$. Furthermore, both interaction potentials for lanthanide atoms have similar contribution to the scattering cross section, in contrast to polar molecules at strong electric fields, where dipolar interactions typically dominate~\cite{Bohn2009}.
 
 There are two competing symmetries encapsulated in the Hamiltonian \eqref{hamiltonian}. At large distances between atoms, the trapping potential with cylindrical symmetry dominates. At short distances, this symmetry is broken by the spherical van der Waals interaction. Furthermore, for \(\theta_F \neq 0\), the cylindrical symmetry is also not respected by the dipole-dipole term, such that angular momentum \( m_l \) is not a good quantum number. The interplay between competing symmetries may give rise to rich physics of confinement-induced resonances. In this work we use the Hamiltonian \eqref{hamiltonian} as a starting point for the dimensional reduction. Our goal is to relate an effective 1D model to the solutions of the full 3D one.
 
\subsection{Multichannel calculations}

Let us briefly discuss the numerical methods for solving the trapped 3D system. For an arbitrary potential, the wavefunction can be expanded in terms of partial waves

\begin{equation}
    \Psi(r,\theta,\varphi) = \frac{1}{r} \sum_{\ell,m_{\ell}} R_{\ell,m_{\ell}}(r) Y_{\ell,m_{\ell}}(\theta,\varphi),
\end{equation}

where \( R_{\ell,m_{\ell}}(r) \) are the radial wavefunctions, and \( Y_{\ell,m_{\ell}}(\theta,\varphi) \) are the spherical harmonics.

Substituting this expansion into the Schr\"odinger equation transforms it into an infinite set of coupled ordinary differential equations.

\begin{equation}
   \left[ -\frac{d^2}{dr^2}  + V_{\ell', \ell,m_{\ell'},m_{\ell}}^{ad}(r) \right]R_{\ell,m_{\ell}}(r) = E \cdot  R_{\ell,m_{\ell}}(r),
\end{equation}
where the coupling potential terms are given by:
\begin{equation}
   V_{\ell', \ell,m_{\ell'},m_{\ell}}^{ad}(r) = \bra{\ell',m_{\ell'}} \hat{V} \ket{\ell, m_{\ell}}.
   \label{eq:pot-adiab}
\end{equation}

The adiabatic potentials can be obtained by diagonalizing the matrix representation of the potential in spherical coordinate basis at fixed distance~\cite{Quemener_2015}. 
Exemplary adiabatic potential curves for the case of Dysprosium at \(\theta_F = 0\) are shown in Fig. \ref{fig:Rysunek 1}E. To account for the cylindrical symmetry at large distances it is crucial to consider large number of channels. 

\begin{figure*}[t]
    \centering   \includegraphics[width=\linewidth]{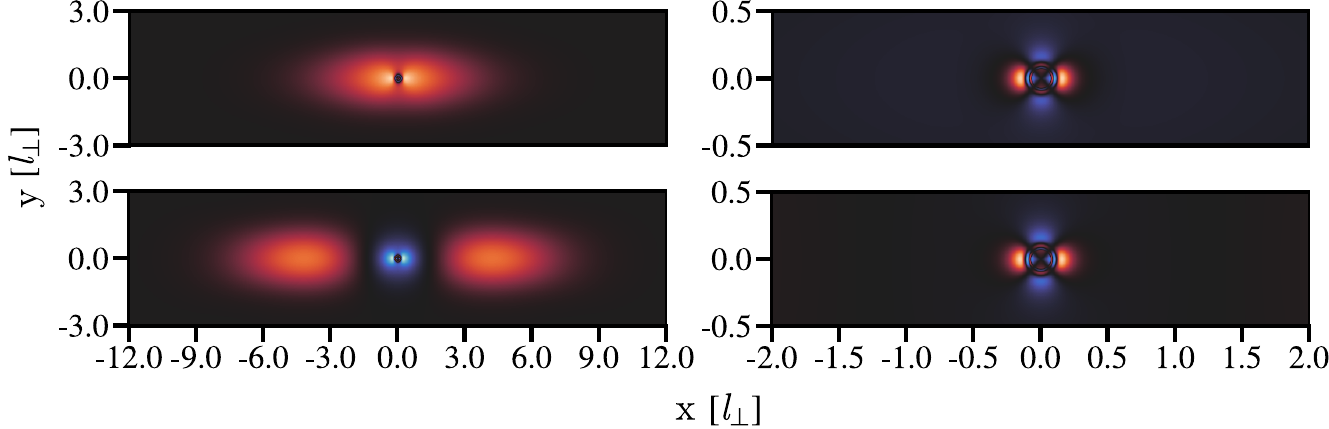}
    \caption{Cross sections of wavefunction at $\varphi=0$ for bound states of two Erbium particles in the harmonic trap at energies $\text{E}=10 \text{E}_{\omega_ \parallel}$ (top) and $\text{E}=12 \text{E}_{\omega_ \parallel}$ (bottom). The $s$-wave states are to good extent bound by the fundamental transverse mode. The $d$-waves states are highly localized and bound mostly by dipolar attraction rather than harmonic trapping.}
    \label{fig:fig3}
\end{figure*}

In order to obtain the energy spectrum we need to specify the boundary conditions at $r\to 0$. As discussed, the exact form of the short-ranged interactions depends on the atomic species and their electronic and hyperfine structure. For complex atoms such as Dysprosium and Erbium, high spin and open electronic $f$-shell makes the problem rather involved. Here we simplify the problem by utilizing Quantum Defect Theory (QDT) to accurately describe the bound states of the 3D system. The method relies on the dominance of the isotropic van der Waals interaction at short range, such that close to the origin we may impose a zero energy solution of the pure van der Waals interaction, parametrized by the short-range phase $\varphi$. The corresponding scattering length $a_{\text{vdW}}$ is then given by~\cite{Gao1998}
\( a_{\text{vdW}}(\varphi) = \frac{\pi}{8} (1 + \tan \varphi)/\Gamma^2\left( \frac{5}{4} \right).\) To correctly relate this to the actual scattering length of the full potential \( a_{3 \text{D}} \), we perform 3D scattering calculations in free space. The conversion between \( a_{\text{vdW}} \) and \( a_{3 \text{D}} \), as well as the solutions of the system Schrödinger equation for the trapped system, are determined using multichannel Numerov method~\cite{Johnson1977,Johnson1978}. An exemplary spectrum of the three-dimensional system is shown in Figure~\ref{fig:fig2}. One immediate observation is the presence of multiple molecular bound states, which are all coupled to the lowest trap level. This is the result of the anisotropy of the dipolar interaction which couples all even partial waves. The broader resonance is associated with a bound state of mostly $s$-wave character, while the narrow one is predominantly $d$-wave. Plotting wavefunctions for different energies reveals the structure of bound states at corresponding resonances (in~Fig.~\ref{fig:fig2}B). Close to these resonances, 3D scattering does not have universal dipolar character~\cite{Bohn2009}, i.e. the energy levels do not depend only on the 3D scattering length and dipole strength. This leads to two distinct branches of states visible in~Fig.~\ref{fig:fig2}B. Consequently, the mapping onto a 1D model will depend on the specific resonance used in experiment.


\section{Dimensional reduction}

We now discuss the procedure of dimensional reduction. In our approach, we first find numerically the full energy spectrum of the trapped system, and then compare the low energy states with the spectrum obtained for a purely 1D model, trying to find the appropriate form of the effective 1D potential. Here for simplicity we assume the symmetric scenario \(\theta_F = 0\). As a result, the angular momentum projection on the trap axis is conserved and we set it to $m=0$. The energy levels of the transverse 2D trap can be labelled as $E_\perp=\hbar\omega_\perp(2k+m+1)$, and the adiabatic potentials converge to these values as seen in Fig.~\ref{fig:Rysunek 1}E,F. Furthermore, we focus on the branch of states in the vicinity of the predominantly s-wave molecular bound state. The first problem is to fit the form of the function $g^{\rm (1D)}_{\rm dd}U_{\rm dip}(x)$. Previous works~\cite{Sinha2007,Deuretzbacher2010} assumed that at low energies, the particles should occupy the ground state of the transverse harmonic confinement. The transverse degrees of freedom can then be integrated out completely, leading to the following regularized form of the effective dipole-dipole interaction potential~\cite{Giovanazzi2004Nov,Deuretzbacher2010}

\begin{align}
   \label{Erfc}
   &U_{\mathrm{dip}}^{\rm approx}(u) \propto -[1 + 3 \cos(2\theta_F)] \\ &\times \left\{ \left[ -2|u| + \sqrt{\pi} (1 + 2u^2) e^{u^2} \operatorname{erfc}( |u| ) \right] - \frac{4}{3} \delta(u) \right\},
\end{align}

with \(u = x/l_{\perp}\). This expression can be interpreted as the effective interaction potential between two particles occupying the ground transverse mode of the harmonic trap. However, this approximation implicitly assumes cylindrical symmetry, which is only valid for large interatomic separations. More proper description at short distances requires considering the coupling between higher-order harmonic oscillator modes, which is particularly important for repulsive dipole orientation~\cite{Guan2014}. A~potentially more accurate choice would be to employ the lowest adiabatic potential resulting from diagonalization of the dipole-dipole interaction matrix represented in the basis of transverse harmonic oscillator states, shown in Fig.~\ref{fig:Rysunek 1}F. To obtain it, one must first construct the matrix

\begin{equation}
    V_{ij}(x) = \delta_{ij}E_i + \iint \phi_i^\perp(y,z)^* V_\mathrm{dd}(x, y, z) \phi_j^\perp(y,z) dydz
\end{equation}
where $\phi_i^\perp$ represents all possible orbitals in the $YZ$ plane, which are given by associated Laguerre polynomials, and the
indices $i,j$ enumerate these orbitals. For a fixed $x$, the sorted eigenvalues
of this matrix provide the successive adiabatic potentials at a distance $x$
between the particles. Note that for dipolar interaction each integral can be computed analytically, and results in an expressions similar to Eq.~\eqref{Erfc} but with growing length, and a Dirac delta appears in every term with the same coefficient in front.

As shown in Fig.~\ref{fig:Rysunek 1}F, the lowest adiabatic potential obtained this way indeed deviates from Eq.\eqref{Erfc} already at distances of the order of $0.5 \ l_{\perp}$, increasing the depth of the potential, and so possibly shifting the positions of resonances. Furthermore, as illustrated in Fig.~\ref{fig:Rysunek 1}C, the resulting lowest adiabatic mode differs from the first transverse mode. This is further elaborated in Fig.~\ref{fig:fig3} showing 2D cross sections of the 3D wavefunctions of for two representative energies. 
Note that as the 3D interaction diverges at the origin, the lowest adiabatic potential gets deeper with the increasing number of included channels. The physical parameter of this model is the bound state energy.  
Interestingly, we found that for attractive dipole orientation satisfactory agreement between the 3D and 1D spectra can be obtained already at the level of the single mode approximation. This is illustrated in Fig.~\ref{fig:fig3}A. 
The remaining task is to find the short-range coupling constant $g_{1 \text{D}}$. In the absence of dipolar interactions ($g_{dd}^{(3D)} = 0$), the coupling strength can be expressed as~\cite{Olshanii1998Aug}
\begin{align}
    g^{\text{(1D)}}_{\text{dd}} &= -\frac{2 \hbar^2}{m a_{\text{1D}}}
    = \frac{2 \hbar^2 a_{3 \text{D}}}{m l_{\perp}^2} \cdot \frac{1}{1 - C \, a_{3 \text{D}}/l_{\perp}}.
    \label{Olshani}
\end{align}

\begin{figure}[h!]
    \centering
    \includegraphics[width=\linewidth]{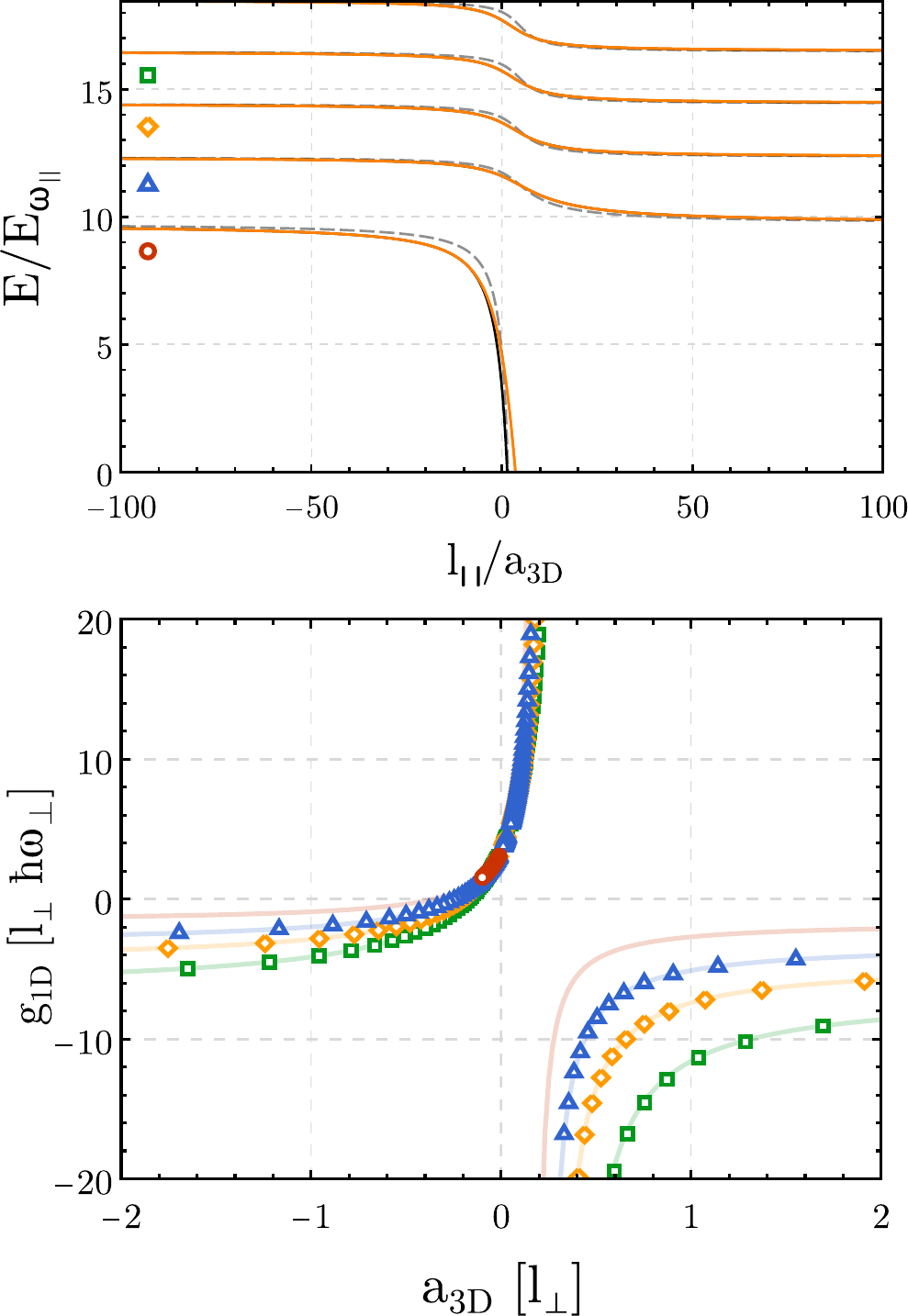}
    \caption{(upper) Energy of two-particle states as a function of $1/a_{3\text{D}}$ for different models: 
        (\raisebox{0.25ex}{\includegraphics[height=0.6ex]{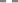}}) 1D model with an effective potential where the first harmonic mode is integrated out, 
        (\raisebox{0.25ex}{\includegraphics[height=0.6ex]{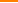}}) full 3D calculation, and 
        (\raisebox{0.35ex}{\includegraphics[height=0.6ex]{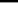}}) renormalized 1D interaction with a 3D correction. 
        After correction, the 1D calculation matches the full 3D results.
        (lower) Relation between $a_{3\text{D}}$ and $g_{1\text{D}}$ for different modes of a cigar-shaped trap:
        \includegraphics[height=1.5ex]{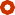} - first, 
        \includegraphics[height=1.5ex]{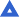} - second, 
        \includegraphics[height=1.5ex]{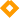} - third, and 
        \includegraphics[height=1.5ex]{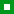} - fourth energy curves (marked with corresponding symbols in Fig.~\ref{fig:fig3}).
        Solid lines represent fits to formula~\eqref{Olshani} with variable position and strength.
        }
    \label{fig:FIG3}
\end{figure}

Here, $C = |\zeta(1/2)|/\sqrt{2} = 1.03263$, and $l_{\perp}$ represents the characteristic transverse harmonic trap length. The key question is whether and to what extent the dipole-dipole interaction affects the applicability of Eq.~\eqref{Olshani}. As the long-range part has been settled, the mapping now becomes straightforward and does not require using the formula~\eqref{Olshani}.
The key result is that with the use of proper regularization of the dipolar potential, the mapping $ (a_{3 \text{D}} , g_{1 \text D})$ follows the functional form of Eq.~\eqref{Olshani} with modified resonance position and strength. Furthermore, we find significant finite-energy corrections, as each branch of the energy spectrum leads to a slightly different fit, shown in Fig.~\ref{fig:fig3}B. This is expected, as the interaction and trapping length scales compete with each other~\cite{Bolda2002}.

Exact 3D calculations thus show that in the regime of tight confinement with strong attractive forces, the confinement-induced resonance (CIR) is shifted, and coupling between higher partial waves becomes significant. Consequently, the energy spectrum exhibits nonuniversal behavior with respect to the 1D short-range parameter $a_{1D}$. The mapping between the 3D scattering length $a_{3 \text{D}}$ and the 1D coupling constant $g_{1 \text{D}}$ is therefore dependent on the specific atom and the Feshbach resonance used to vary $a_{3 \text{D}}$

\subsection{Other system geometries}
\begin{figure}[!h]
    \centering
    \includegraphics[width=0.9\linewidth]{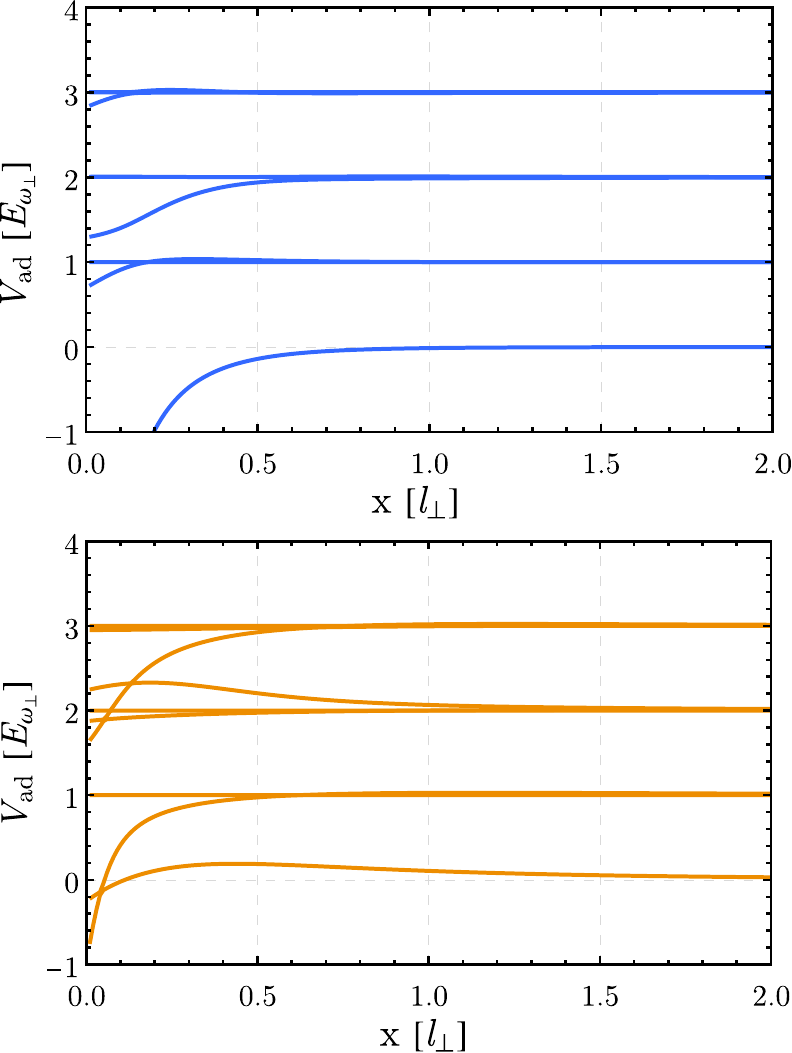}
    \caption{Analysis of the 1D system with dipole-dipole interaction (upper) at the “magic angle” $\theta_F \approx 55 \degree$. The adiabatic potentials diagonalized in the basis of harmonic oscillator modes, (lower) with dipole orientation perpendicular to the trap axis.} 
    \label{fig:FIG4}
\end{figure}

We have by now analyzed the impact of the dipolar potential for the specific case of dipoles oriented parallel to the trap axis, where the interaction is predominantly attractive. A natural question arises as to whether corrections persist when the dipole moments are oriented at other angles, including the “magic angle”, where the dipole-dipole interaction is expected to vanish in the 1D model as predicted by Eq.~\eqref{Erfc}. To address this scenario, we studied the adiabatic potentials in the harmonic oscillator basis, which are presented in Fig.~\ref{fig:FIG4}A for the magic case and ~\ref{fig:FIG4}B for perpendicular dipole orientation, which leads to asymptotically repulsive interaction studied before by~\cite{Guan2014}. Note that now the trap axis is not aligned with the dipole, leading to couplings between channels with different $m$. For identical bosons, we are interested in symmetric wavefunctions, which can be realized with both $m$ parities. Based on previous results, it is now not surprising that the lowest adiabatic potential differs from zero even for the magic angle case. In fact, it has a range comparable to the trap lengthscale and decays at large distance approximately as $x^{-6}$. While this is a short-range potential with a rather weak amplitude, it can be expected to shift $a_{1D}$ from the value predicted by Eq.~\eqref{Olshani}. This shift should be particularly pronounced for weak effective interactions. For the repulsive dipole configuration, the lowest adiabatic potential also becomes attractive at short range, shifting the confinement-induced resonance position. 


\section{Discussion and Conclusions}
Our results demonstrate that a full 3D treatment is necessary to capture the subtle influence of long-range dipole-dipole interactions, especially at short interatomic distances where coupling to excited transverse modes becomes significant. In contrast to the widely used approximations that neglect these couplings, the regularized effective 1D potential accurately reproduces the energy levels observed in full 3D calculations, thus providing a more reliable description of the low-energy physics. These effects are the most significant in the regime of weak contact interactions, where the attractive dipolar interaction dominates. However, in the seemingly opposite case of the “magic angle,” where dipolar interactions are expected to vanish in 1D models, we find that corrections persist due to residual coupling to higher transverse modes. This leads to a renormalization of the effective 1D parameters, which is essential for accurate interpretation of experimental results.

The 3D nature of the interaction becomes particularly important when exploring higher energies by exciting the system or increasing the density, where the effective range of the interaction impacts the coupling strength and shifts the confinement-induced resonance.

Our findings thus directly impact current research on confined dipolar systems and provide a robust framework for analyzing strongly interacting dipolar gases. The combined use of the corrected effective 1D potential and the exact mapping between 3D and 1D scattering parameters accurately describes the system outside of confinement-induced resonance (CIR) regimes, where nonuniversal effects prevail. 

We suggest using the developed framework as a reliable foundation for extending the analysis to molecular dipolar systems~\cite{Bigagli2024Jul}. Correct treatment of dipole-dipole interactions might be crucial in theoretical analysis of quantum phases highly sensitive to the scattering parameters such as the quantum droplets~\cite{Schmidt2022,Kopycinski2023,Langen2025,Lebek2025}.


{\it Acknowledgements.} The authors would like to thank Zbigniew Idziaszek, Piotr Kulik, and Gregory Astrakharchik for useful discussions. M.Z. and K.J. were supported by the National Science Centre of Poland under Grant No. 2020/37/B/ST2/00486. K.P. acknowledges support from the (Polish) National
Science Center Grant No. 2019/34/E/ST2/00289.

\bibliography{refs.bib}

\onecolumngrid
\appendix

\end{document}